\documentclass[11pt, letterpaper]{article}
\usepackage[utf8]{inputenc}
\usepackage{amsmath,amssymb}
\usepackage{graphicx}
\usepackage{subfig}
\usepackage{here}
\usepackage{float}
\usepackage{multirow}
\usepackage[margin=0.75in]{geometry}
\usepackage{color}
\usepackage{ulem}
\usepackage{authblk}
\usepackage{siunitx}
\usepackage{booktabs}
\DeclareMathAlphabet{\mathpzc}{OT1}{pzc}{m}{it}
\newcommand{\vev}[1]{\langle\Omega|#1|\Omega\rangle}

\newcommand{\lsr}{\mathcal{R}_{0}}
\newcommand{\hypgeom}[2]{{}_{#1}F_{#2}}
\newcommand{\double}[2]{(#1,\,#2)}
\newcommand{\dif}[1]{d #1}
\newcommand{\glueFourD}{\big\langle \alpha G^{2} \big\rangle}
\newcommand{\glueSixD}{\big\langle g^{3} G^{3} \big\rangle}

\newcommand{\rom}[1]{\uppercase\expandafter{\romannumeral #1\relax}}

\title{Meson-Hybrid Mixing in $J^{PC}=1^{++}$ Heavy Quarkonium from QCD Sum-Rules}
\author[1]{A.\ Palameta}
\author[2]{D.\ Harnett}
\author[1]{T.G.\ Steele}
\affil[1]{Department of Physics and Engineering Physics\\ University of Saskatchewan\\           Saskatoon, SK, S7N 5E2, Canada}
\affil[2]{Department of Physics\\ University of the Fraser Valley\\ Abbotsford, BC, V2S 7M8, Canada}
\begin{document}
\maketitle
\begin{abstract}
\noindent We explore conventional meson-hybrid mixing in $J^{PC}=1^{++}$ heavy quarkonium using QCD Laplace sum-rules. We calculate the cross-correlator between a heavy conventional meson current and heavy hybrid current within the operator product expansion, including terms proportional to the four- and six-dimensional gluon condensates and the six-dimensional quark condensate. Using experimentally determined hadron masses, we construct models of the $1^{++}$ charmonium and bottomonium mass spectra. These models are used to investigate which resonances couple to both currents and thus exhibit 
conventional meson-hybrid mixing. 
In the charmonium sector, we find almost no conventional meson-hybrid mixing in the $\chi_{c1}(1P)$, minimal mixing in the $X(3872)$, and significant mixing in both the
$X(4140)$ and $X(4274)$.
In the bottomonium sector, we find minimal conventional meson-hybrid mixing in the $\chi_{b1}(1P)$
and significant mixing in both the $\chi_{b1}(2P)$ and $\chi_{b1}(3P)$.
\end{abstract}

\section{Introduction}\label{I}

Hybrids are hadrons which consist of a quark-antiquark pair and exhibit explicit gluon degrees of freedom. 
Hybrids are allowed by QCD as they are colour singlets; however, they have not yet been definitively experimentally identified~\cite{Meyer:2015eta}. 

Hybrids can be classified by $J^{PC}$, quantum numbers that can be separated into two categories, non-exotic and exotic, depending on whether the quantum numbers are accessible to conventional (quark-antiquark) mesons or not. Hybrids with exotic $J^{PC}$ would not be able to quantum mechanically mix with conventional mesons;  however, hybrids with non-exotic quantum numbers can potentially mix with conventional mesons. This mixing would result in hadrons that are superpositions of both conventional meson and hybrid.

In this article, we extend our work from~\cite{Palameta:2017ols} on 
vector (i.e., $1^{--}$) conventional meson-hybrid mixing
to axial vector (i.e., $1^{++}$) charmonium ($c\overline{c}$) and bottomonium ($b\overline{b}$). 
Of particular interest in the charmonium sector is the 
$X(3872)$~\cite{Choi:2003ue,Olive:2016xmw}, 
the first of the XYZ 
resonances~\cite{Swanson:2006st,Zhu:2007wz,Godfrey:2008nc,Nielsen:2009uh,Brambilla:2010cs},
a collection of charmonium-like hadrons many of which are not easily 
accommodated by the constituent quark model.
The $X(3872)$ has been studied in the context of conventional meson-tetraquark 
mixing~\cite{Matheus:2009vq} as well as tetraquark-hybrid mixing~\cite{ChenJinKleivEtAl2013} (see also~\cite{Padmanath:2015era,Kang:2016jxw} for other approaches to mixing).
Our analysis complements these two by considering conventional meson-hybrid mixing.
At present, the $1^{++}$ channel is the only channel other then the $1^{--}$ with enough experimentally observed resonances to allow for the
multi-resonance analysis methods of~\cite{Palameta:2017ols}.

We use the operator product expansion (OPE)~\cite{Wilson:1969zs} to compute the cross-correlator between a heavy conventional meson current and a heavy hybrid current. In this calculation we include leading-order (LO) contributions from perturbation theory and non-perturbative corrections proportional to the four-dimensional (4d) and 6d gluon condensates as well as the 6d quark condensate. Then, using QCD Laplace sum-rules (LSRs)~\cite{Shifman:1978bx,Shifman:1978by,Reinders:1984sr,narisonbook:2004}, we analyze several single and multi-resonance models of the $1^{++}$ charmonium and bottomonium mass spectra. These models take known resonance masses as inputs and allow us to probe the resonances to determine whether they couple to both the conventional meson current and the hybrid current. Resonances which couple to both currents are considered to be quantum mechanical mixtures of conventional meson and hybrid. The QCD sum-rules methodology has been applied to hadron mixing in a number of systems~\cite{Palameta:2017ols,Narison:1984bv,Harnett:2008cw,Chen:2013pya,Ho:2016owu}.

We find that multi-resonance models which include excited states in addition to the ground state lead to a significant improvement in agreement between QCD and experiment when compared to single resonance models. We show explicitly that the higher mass states make numerically significant contributions to the LSRs despite the LSR's exponential suppression of such resonances. 
In the charmonium sector, we find very little 
conventional meson-hybrid mixing in the $\chi_{c1}(1P)$, 
minimal mixing in the $X(3872)$, 
and large mixing in both the $X(4140)$ and the $X(4274)$. 
In the bottomonium sector, we find minimal conventional meson-hybrid 
mixing in the $\chi_{b1}(1P)$
and large mixing in both the $\chi_{b1}(2P)$ and $\chi_{b1}(3P)$.

\section{The Correlator}\label{II}
For the conventional meson current
\begin{equation}\label{CurMes}
  j_{\mu}^{(\mathrm{m})} = \overline{Q}\gamma_{\mu}\gamma^{5} Q
\end{equation}
and the hybrid current~\cite{GovaertsReindersWeyers1985}
\begin{equation}\label{CurHyb}
  j_{\nu}^{(\mathrm{h})}=  \frac{g_{s}}{2}\overline{Q}\gamma^{\rho}\lambda^{a} \widetilde{G}^{a}_{\nu\rho} Q
\end{equation}
where $Q$ is a heavy quark (i.e., charm or bottom) field and
\begin{equation}\label{dualFieldStrength}
  \widetilde{G}^{a}_{\nu\rho} = \frac{1}{2}\epsilon_{\nu\rho\omega\zeta}G^{a}_{\omega\zeta}
\end{equation}
is the dual gluon field strength tensor, we consider the cross-correlator
\begin{align}
  \Pi_{\mu\nu}(q) &= i\!\int d^{4}\!x \;e^{iq\cdot x} 
    \vev{\tau\, j_{\mu}^{(\text{m})}(x)\; j_{\nu}^{(\text{h})}(0)}
    \label{CorFn}\\
   &= \frac{q_{\mu}q_{\nu}}{q^2}\Pi_0(q^2) + \left(\frac{q_{\mu}q_{\nu}}{q^2}-g_{\mu\nu}\right)\Pi_1(q^2).
   \label{CorFnProj}
\end{align}
In~(\ref{CorFnProj}), the function $\Pi_0(q^{2})$ probes spin-0 states and $\Pi_1(q^{2})$ probes spin-1 states. 
We focus on $\Pi_1(q^2)$ as we are interested in probing $1^{++}$ states.

We evaluate the cross-correlator~(\ref{CorFn}) within the OPE where perturbation theory is supplemented by non-perturbative corrections. Each of these non-perturbative corrections is the product of a perturbatively computed Wilson coefficient and a QCD condensate. We include terms proportional to the 4d and 6d gluon condensates and the 6d quark condensate defined respectively 
as follows:
\begin{equation} \label{fourDcond}
\big\langle \alpha G^{2} \big\rangle = \alpha_{s} \big\langle \! \colon \! G_{\omega\phi}^{a} G_{\omega\phi}^{a} \! \colon\! \big\rangle
\end{equation}
\begin{equation} \label{sixDcond}
\begin{aligned}
\big\langle g^{3}G^{3}\big\rangle = g_{s}^{3} f^{abc} \big\langle \! \colon \! G^{a}_{\omega\zeta} \; G^{b}_{\zeta\rho} G^{c}_{\rho\omega} \! \colon\! \big\rangle
\end{aligned}
\end{equation}
\begin{equation} \label{sixDQcondTr}
\big\langle J^2 \big\rangle =
\mathrm{Tr} \big(\big\langle \! \colon\! J_{\nu} J_{\nu} \! \colon\! \big\rangle \big)
\end{equation}
with
\begin{equation} \label{jDef}
J_{\nu} 
= \frac{-i g_s^2}{4} \lambda^a
\sum_{A} \overline{q}^A \lambda^a \gamma_{\nu} q^A \end{equation}
where, in~(\ref{jDef}), $q$ is a light quark (i.e., up, down, or strange) field
and the sum is over flavours.
We use the vacuum saturation hypothesis~\cite{Shifman:1978bx} to express
$\langle J^2 \rangle$ in terms of the 3d quark condensate
\begin{equation} \label{threeDcond}
\big\langle \overline{q}q \big\rangle = \big\langle \! \colon \! \overline{q}^{\alpha}_{i} q^{\alpha}_{i} \! \colon\! \big\rangle
\end{equation}
resulting in
\begin{equation} \label{sixDQcond}
\big\langle J^2 \big\rangle
= \frac{2}{3} \, \kappa \, g_s^4 \big\langle \overline{q}q \big\rangle^2
\end{equation}
where $\kappa$ quantifies deviations from vacuum saturation. 
As in~\cite{Palameta:2017ols}, we set $\kappa=2$~\cite{narisonbook:2004}.

\begin{figure}
\centering
\begin{tabular}{ccc}
\includegraphics[width=50mm]{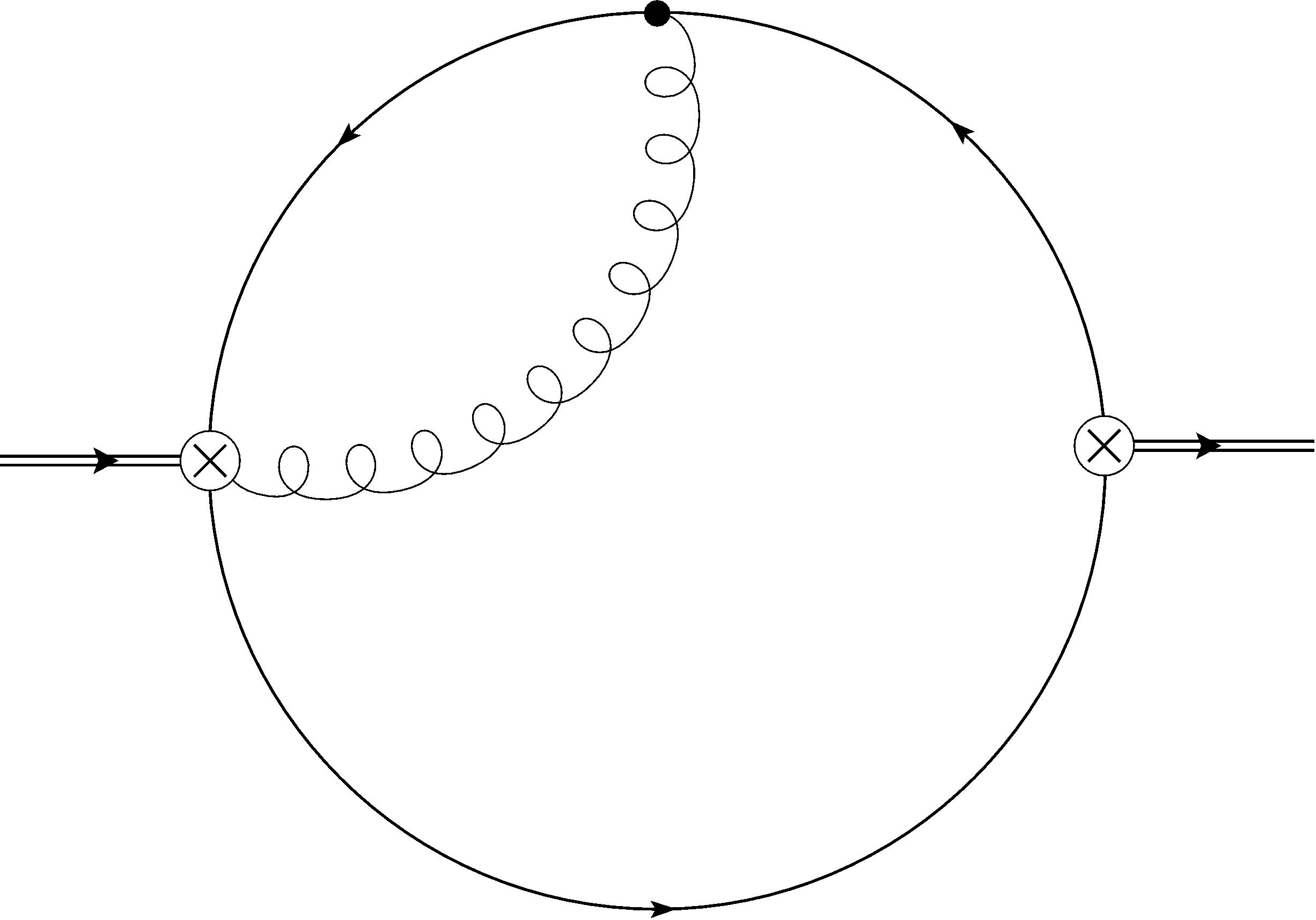} & \includegraphics[width=50mm]{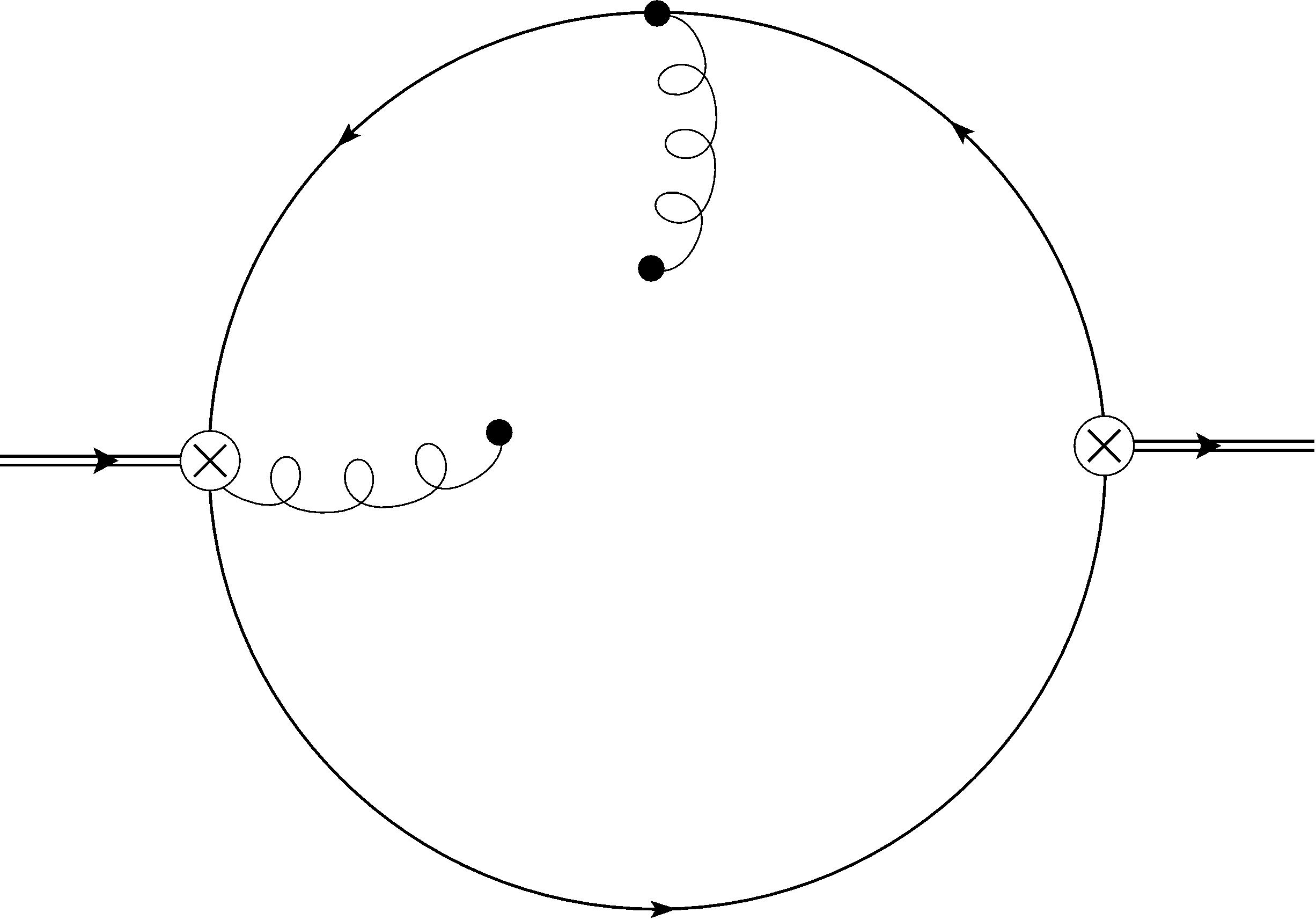} & \includegraphics[width=50mm]{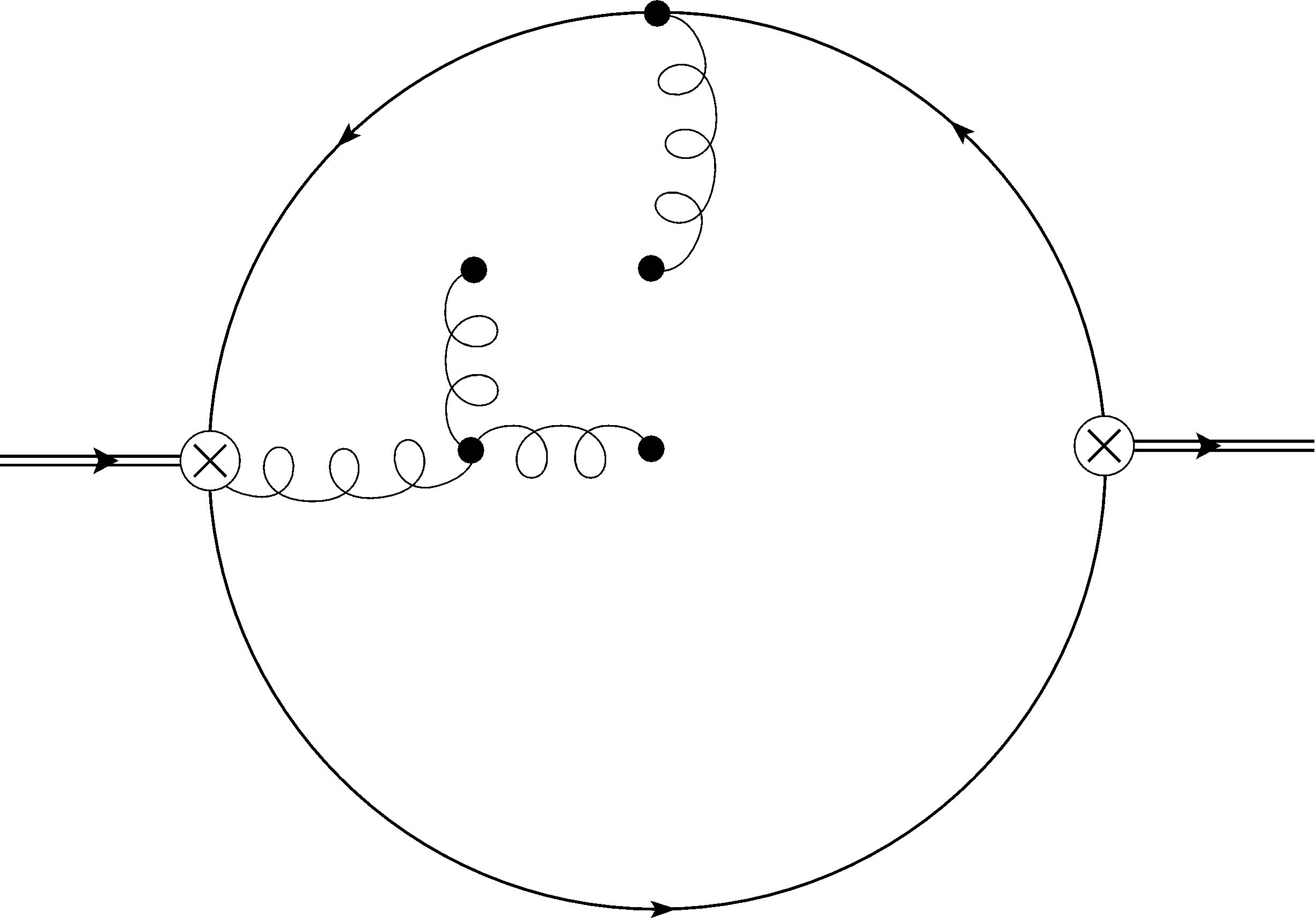}\\
Diagram \rom{1} & Diagram \rom{2} & Diagram \rom{3} \\[15pt]
\includegraphics[width=50mm]{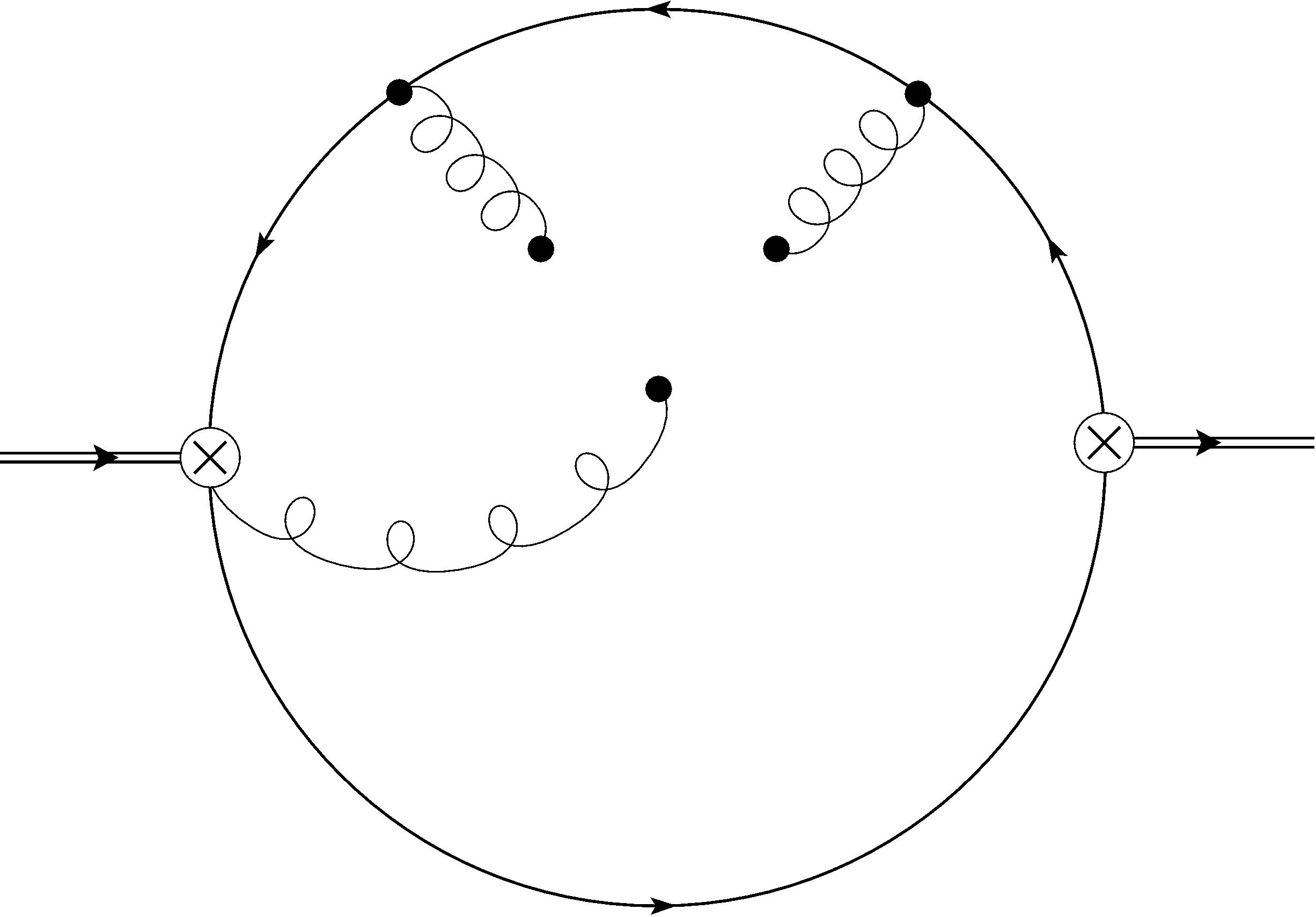} & \includegraphics[width=50mm]{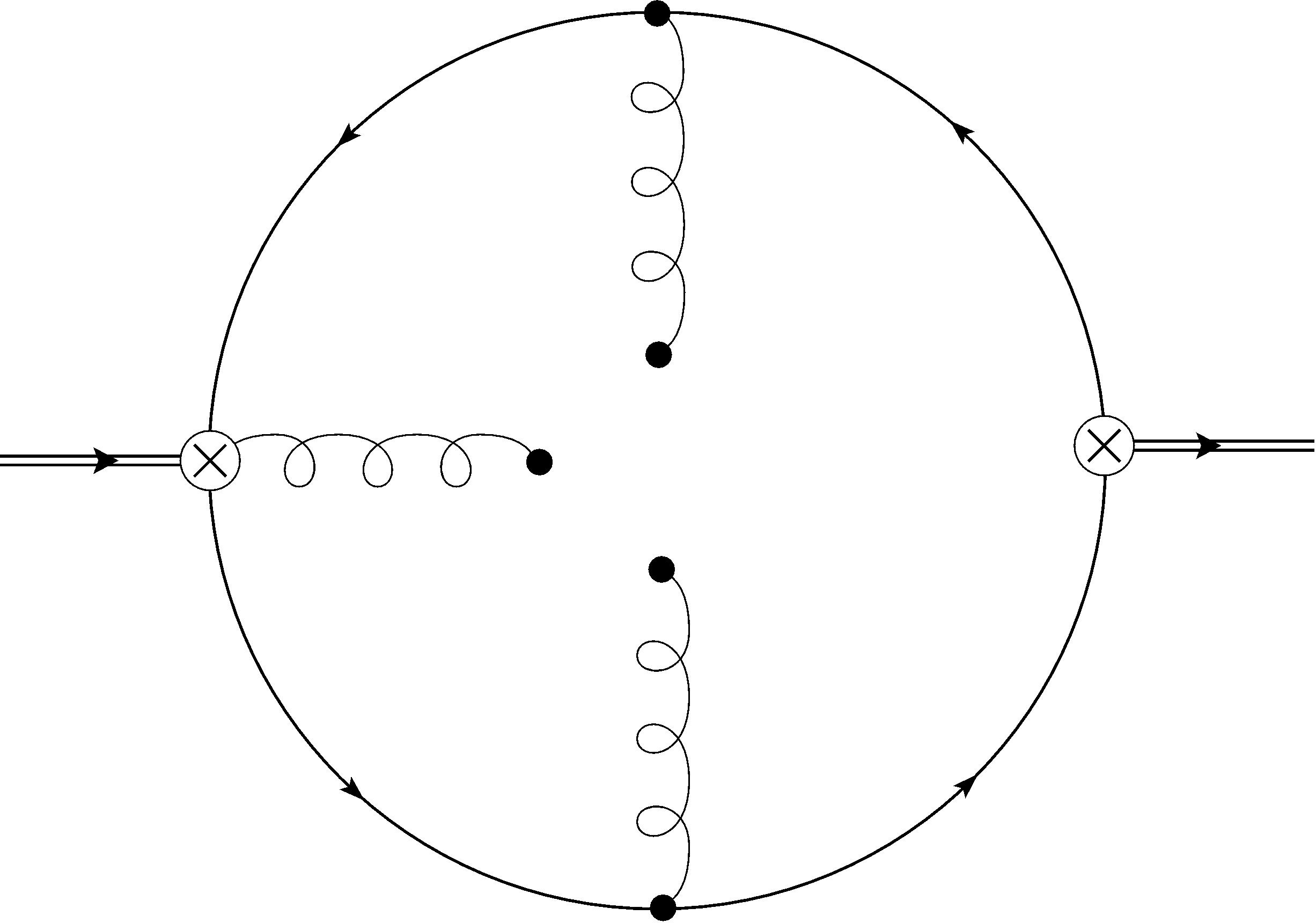} & \includegraphics[width=50mm]{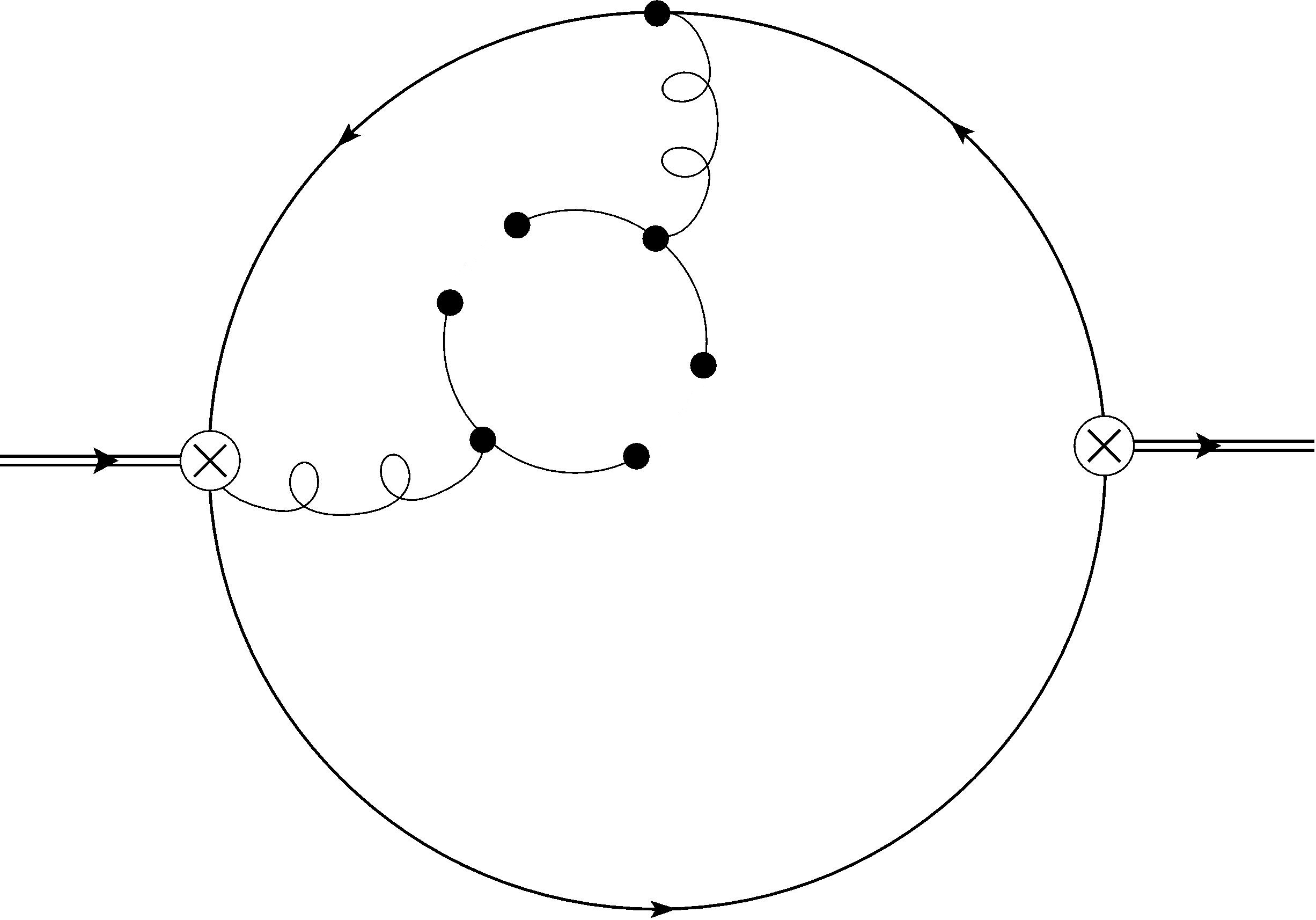}\\
Diagram \rom{4} & Diagram \rom{5} & Diagram \rom{6} \\[15pt]
\end{tabular}
\caption{Feynman diagrams that contribute to the cross-correlator~(\ref{CorFn}) at LO.}
\label{fig01}
\end{figure}

The diagrams that contribute to~(\ref{CorFn}) at LO are given in Figure~\ref{fig01}. 
Each diagram is multiplied by two to account for additional diagrams in which the
quark lines run in the opposite directions.
Diagram~\rom{4} gets another factor of two due to 
symmetry under exchange of the two interaction vertices.
We calculate Wilson coefficients via fixed-point gauge methods~\cite{PascualTarrach1984,BaganAhmadyEliasEtAl1994}, and divergent integrals are dealt with through dimensional regularization in $D=4+2\epsilon$ dimensions
at $\overline{\text{MS}}$-scale $\mu$.
The Mathematica package TARCER~\cite{MertigScharf1998} which implements the recurrence relations of~\cite{Tarasov1996,Tarasov1997} is used to express results in terms of known master integrals including those of~\cite{BoosDavydychev1991,BroadhurstFleischerTarasov1993}. 
Following~\cite{AkyeampongDelbourgo1973} 
we use the $\gamma^{5}$ convention
\begin{equation} \label{gamma5conv}
\gamma^{5}=\frac{i}{24}\epsilon_{\mu\nu\sigma\rho}\gamma^{\mu}\gamma^{\nu}\gamma^{\sigma}\gamma^{\rho}.
\end{equation}
The OPE computation of $\Pi_1$ from~(\ref{CorFnProj}), denoted $\Pi^{(\text{OPE})}$,
is decomposed as 
\begin{equation} \label{fullCorForm}
  \Pi^{(\text{OPE})} = \Pi^{(\mathrm{\rom{1}})} + \Pi^{(\mathrm{\rom{2}})} + \Pi^{(\mathrm{\rom{3}})} + \Pi^{(\mathrm{\rom{4}})} + \Pi^{(\mathrm{\rom{5}})} + \Pi^{(\mathrm{\rom{6}})}
\end{equation}
where the superscripts in~(\ref{fullCorForm}) correspond to the labels of the diagrams in Figure~\ref{fig01}. For $\Pi^{(\text{I})}$, the $\epsilon$-dependent result is given by
\begin{multline} \label{pertPreXp}
  \Pi^{(\mathrm{\rom{1}})}(z ; \epsilon) = \frac{\alpha_s \; m^{4(1+\epsilon)} (1+\epsilon) \Gamma^2(-1 -\epsilon)}{6 \pi^3 z (3+2 \epsilon)^{2} (4+3 \epsilon) (4 \pi)^{2 \epsilon}}\Bigg( \\
  -12 z (\epsilon +1) (2 \epsilon +1) (3 \epsilon +4) \left(z (4 \epsilon +5)-2 \epsilon ^2-7 \epsilon -5\right) \, _2F_1\left(1,-\epsilon ;\frac{3}{2};z\right) \\
  +\left(4 z^2 (\epsilon +1) (2 \epsilon +3) (7 \epsilon +8)+z (2 \epsilon +1) (\epsilon  (4 \epsilon +7)+4) +(\epsilon +2) (2 \epsilon +1) (4 \epsilon +5)\right) \\
 \times \, _3F_2\left(1,-2 \epsilon -1,-\epsilon ;\frac{1}{2}-\epsilon ,\epsilon +2;z\right) \\
 + 2 (z-1) (2 \epsilon +1) (z (\epsilon  (8 \epsilon +19)+12)+(\epsilon +2) (4 \epsilon +5)) \, _3F_2\left(1,-2 \epsilon ,-\epsilon ;\frac{1}{2}-\epsilon ,\epsilon +2;z\right)
  \Bigg)
\end{multline}
where
\begin{equation}
  z=\frac{q^2}{4m^2},
\end{equation}
and all polynomials in $z$ have been omitted as they will not contribute to the LSR. In~(\ref{pertPreXp}), $m$ is a heavy quark mass, $\Gamma$ is the Gamma function, and 
$\hypgeom{p}{q}(\cdots;\cdots;z)$ are generalized hypergeometric functions~\cite{AbramowitzStegun1965}. 
Expanding~(\ref{pertPreXp}) in $\epsilon$, we find
\begin{equation} \label{pertXpNow}
\Pi^{(\mathrm{\rom{1}})}(z) =\frac{20\alpha_s m^4 z(z-1)\:\hypgeom{2}{1}\left(1,1;\frac{5}{2};z\right)}{27\pi^3}\frac{1}{\epsilon}
+\frac{d}{d\epsilon}\Pi^{(\mathrm{\rom{1}})}(z ; \epsilon) \Big|_{\epsilon = 0}.
\end{equation}
We do not include an explicit expression for the derivative term on the right-hand side of~(\ref{pertXpNow}) as it will be replaced by~(\ref{expRenormedPert}) shortly. Expanding the remaining terms from~(\ref{fullCorForm}) in $\epsilon$, we find
\begin{gather}
\Pi^{(\mathrm{\rom{2}})}(z) = \frac{z\Big(3-\hypgeom{2}{1}\big(1,1;\frac{5}{2};z\big)\Big)}{36\pi(z-1)}
\big\langle\alpha G^{2} \big\rangle
\label{expfourd}\\
\Pi^{(\mathrm{\rom{3}})}(z) = \frac{\Big(-3 \left(44 z^2-108 z+73\right)+\left(24 z^3-56 z^2+38 z+3\right)\:\hypgeom{2}{1}\big(1,1;\frac{5}{2};z\big)\Big)}{13824\pi^2 m^2 (z-1)^3}
\big\langle g^{3} G^{3} \big\rangle
\label{expsixd}\\
\Pi^{(\text{\rom{4}})}(z) = \frac{\big\langle g^{3} G^{3} \big\rangle}{13824\pi^2 m^2 (z-1)^2}\Bigg(132 z-183 + \left(-24 z^2+38 z+3\right)\:
\hypgeom{2}{1}\big(1,1;\textstyle{\frac{5}{2}};z\big)\Bigg)
\label{expDiag4}\\
\Pi^{(\text{\rom{5}})}(z) = \frac{\big\langle g^{3} G^{3} \big\rangle}{4608\pi^2 m^2 (z-1)^2}\Bigg(12z-15-(2z-3)\:
\hypgeom{2}{1}\big(1,1;\textstyle{\frac{5}{2}};z\big)\Bigg)
\label{expDiag5}\\
\Pi^{(\text{\rom{6}})}(z) = \frac{4 \alpha_s^2 \big\langle \overline{q}q \big\rangle^2}{243 m^2 (z-1)^3}\Bigg(3 \left(44 z^2-108 z+73\right) - \left(24 z^3-56 z^2+38 z+3\right)\:
\hypgeom{2}{1}\big(1,1;\textstyle{\frac{5}{2}};z\big)\Bigg).
\label{expDiag6}
\end{gather}
 
\begin{figure}
\centering
\begin{tabular}{cc}
\includegraphics[width=50mm]{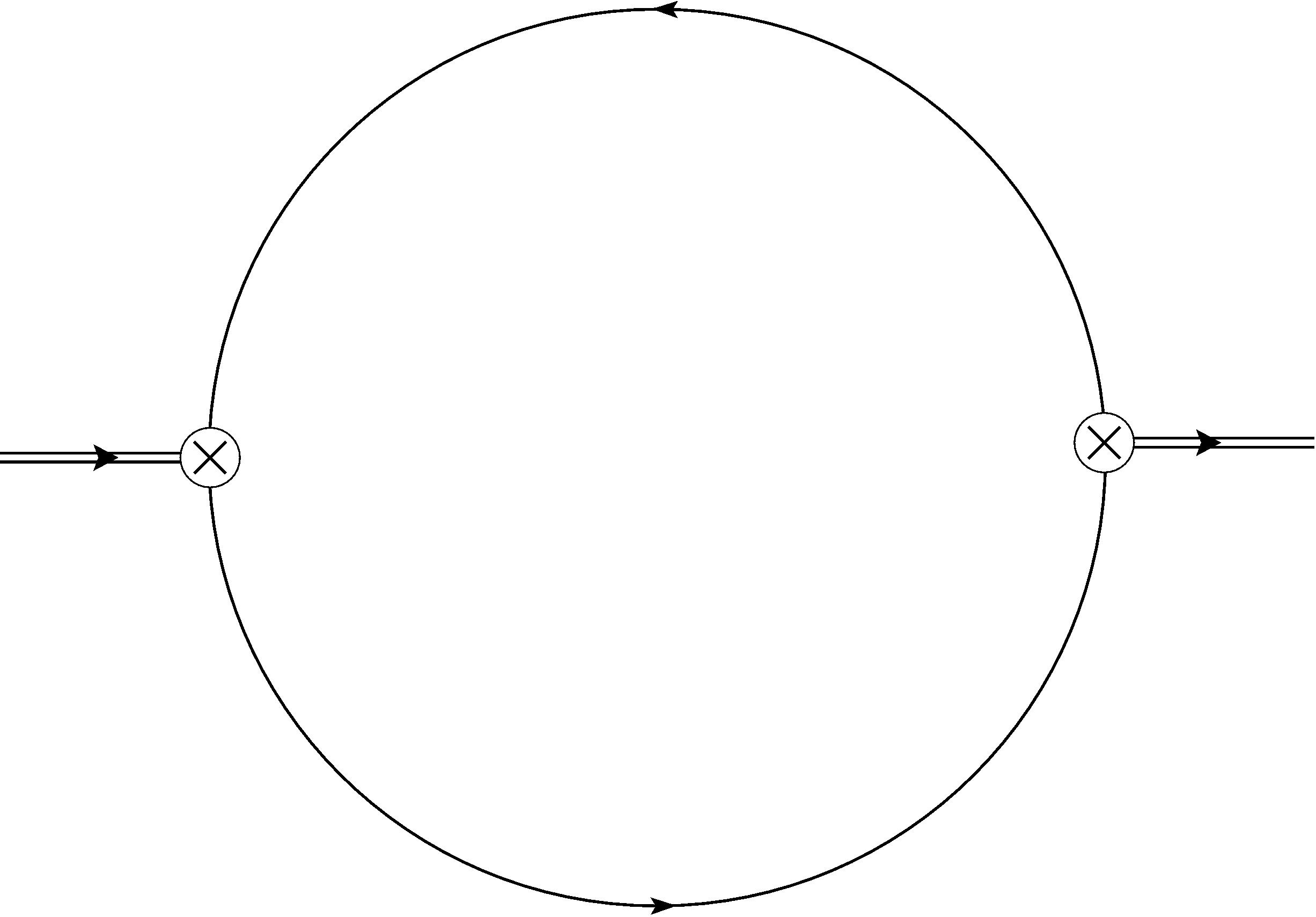} & \includegraphics[width=50mm]{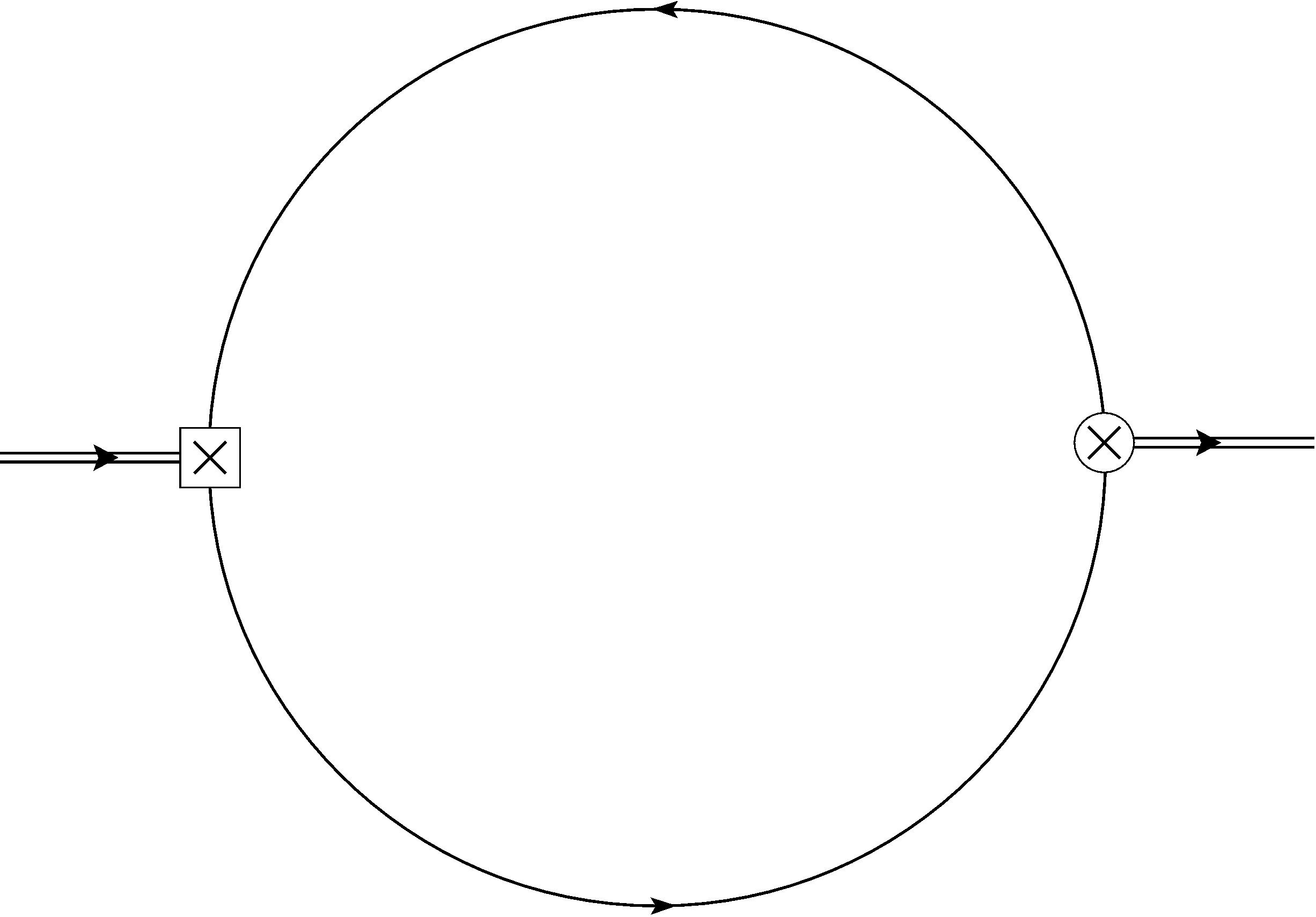} \\
Diagram R\rom{1} & Diagram R\rom{2} \\[15pt]
\end{tabular}
\caption{Renormalization-induced Feynman diagrams resulting from~(\ref{renormBig}). The square insertion denotes the current~(\ref{DCurrent}).}
\label{fig02}
\end{figure}

The perturbative result~(\ref{pertXpNow}) contains a nonlocal divergence. We eliminate this nonlocal divergence through operator mixing under renormalization as 
in~\cite{Palameta:2017ols,Chen:2013pya,Ho:2016owu}. The meson current~(\ref{CurMes}) is renormalization-group (RG) invariant, and so we only need to consider the operator mixing of the hybrid current~(\ref{CurHyb}). 
The only operators that can mix with~(\ref{CurHyb}) 
and possibly generate nonzero contributions to the LO renormalized correlator are
$j^{(m)}_{\nu}$ given in~(\ref{CurMes}) and
\begin{equation} \label{DCurrent}
j_{\nu}^{(c)} = \overline{Q} i \gamma^5 D_{\nu} Q
\end{equation}
where $D_{\nu}=\partial_{\nu}-\frac{i}{2} g_s \lambda^a A^a_{\nu}$ is the covariant
derivative.  
Then, the replacement 
\begin{equation} \label{renormTheLast}
  j_{\nu}^{(\text{h})}  \rightarrow  j_{\nu}^{(\text{h})} 
  + Z_{1} \, j_{\nu}^{(\text{m})} 
  + Z_{2} \, j_{\nu}^{(\text{c})}
\end{equation}
for renormalization constants $Z_1$ and $Z_2$ must result in a perturbative contribution free of nonlocal divergences. Substituting~(\ref{renormTheLast}) into~(\ref{CorFn}) in $D$ dimensions gives
\begin{multline} \label{renormBig}
   i\! \int \!\! d^{D}\!x \; e^{iq\cdot x} \langle\Omega | 
   \tau j_{\mu}^{(\text{m})} j_{\nu}^{(\text{h})} |\Omega\rangle
   \rightarrow 
   i\! \int \!\! d^{D}\!x \; e^{iq\cdot x} \langle\Omega|\tau j_{\mu}^{(\text{m})} 
   j_{\nu}^{(\text{h})} |\Omega\rangle \\
   +i \, Z_{1} \int \!\! d^{D}\!x \; e^{iq\cdot x}\langle\Omega|\tau j_{\mu}^{(\text{m})}
   j_{\nu}^{(\text{m})} |\Omega\rangle 
   +i \, Z_{2} \int\!\! d^{D}\!x \; e^{iq\cdot x} \langle\Omega|\tau j_{\mu}^{(m)}
   j_{\nu}^{(\text{c})} | \Omega \rangle.
\end{multline}
The two terms in~(\ref{renormBig}) containing $Z_1$ and $Z_2$ each generate a renormalization-induced diagram, both of which are shown in Figure~\ref{fig02}. Evaluating these two diagrams and selecting $Z_1$ and $Z_2$ such that the right-hand side of~(\ref{renormBig}) is free of nonlocal divergences, we find
\begin{gather}
Z_{1} = -\frac{10 m^2 \alpha_s}{9 \pi \epsilon} \label{Z1}\\
Z_{2} = -\frac{4 m \alpha_s}{9 \pi \epsilon} \label{Z2}.
\end{gather}
Substituting~(\ref{Z1}) and~(\ref{Z2}) into~(\ref{renormBig}) and expanding in $\epsilon$
gives a renormalized expression
\begin{multline} \label{expRenormedPert}
  \Pi^{(\text{\rom{1}})}(z) = \frac{m^4 \alpha_s}{243 \pi^3} 
  \Bigg(9 (6z^2 -z -5)\:\hypgeom{3}{2}\big(1,1,1;\tfrac{3}{2},3;z\big)
  -z \left(48 z^2+2 z+5\right)\:\hypgeom{3}{2}\big(1,1,2;\tfrac{5}{2},4;z\big)\\
  +9z \left( 20 (z-1) \log \left(\frac{m^2}{\mu ^2}\right) +8z +5 \right) \:\hypgeom{2}{1}\big(1,1;\tfrac{5}{2};z\big) \Bigg)
\end{multline}
where, again, we have omitted polynomials in $z$ as they will not contribute to the LSR.

Finally, collecting~(\ref{expfourd})--(\ref{expDiag6}) and~(\ref{expRenormedPert}) 
and then substituting them into~(\ref{fullCorForm}) gives us the LO expression for $\Pi^{(\text{OPE})}$ up to 6d condensates.

\section{QCD Laplace Sum-Rules}\label{III}
For Euclidean momentum $Q^{2}=-q^{2}>0$, 
the quantity $\Pi_1$ from~(\ref{CorFnProj}) 
satisfies the dispersion relation
\begin{equation}\label{dispersion_relation}
  \Pi(Q^2)=\frac{Q^6}{\pi}\int_{t_0}^{\infty}
  \frac{\mathrm{Im}\Pi(t)}{t^3(t+Q^2)}
  \dif{t} +\cdots
\end{equation}
where $\Pi$ on the left-hand side represents the QCD result $\Pi^{\text{(OPE)}}$
and $\mathrm{Im}\Pi(t)$ on the right-hand side is the hadronic spectral function. 
Equation~(\ref{dispersion_relation}) is a statement of quark-hadron duality and allows us to interpret QCD information contained in the cross-correlator in the context of hadrons. 
In~(\ref{dispersion_relation}), $t_0$ is the hadron production threshold and the $\cdots$ represents unknown subtraction constants (a polynomial in $Q^2$). 
To eliminate these subtraction constants, eliminate local divergences in $\Pi^{\text{(OPE)}}$,
and accentuate the resonance contributions of the hadronic spectral function,
we apply to~(\ref{dispersion_relation}) the Borel transform 
\begin{equation}\label{borel}
  \hat{\mathcal{B}}=\!\lim_{\stackrel{N,Q^2\rightarrow\infty}{\tau=N/Q^2}}
  \!\frac{\big(-Q^2\big)^N}{\Gamma(N)}\bigg(\frac{d}{dQ^2}\bigg)^N
\end{equation}
where $\tau$ is the Borel parameter. This results in the formation of the $0^{\text{th}}$-order LSR~\cite{Shifman:1978bx}
\begin{equation} \label{zeroThLSR}
\lsr(\tau)\equiv\frac{1}{\tau}\hat{\mathcal{B}}\Big\{\Pi(Q^2) \Big\}\\
= \int_{t_0}^{\infty} e^{-t\tau}\frac{1}{\pi}\mathrm{Im}\Pi(t)\dif{t}.
\end{equation}
We then introduce a ``resonance(s) plus continuum'' model
\begin{equation} \label{ResCont}
\frac{1}{\pi}\mathrm{Im}\Pi(t)
    \to \rho^{\text{(had)}}(t)+\frac{1}{\pi}\mathrm{Im}\Pi^{\text{(OPE)}}(t)\theta(t-s_0)
\end{equation}
where $\rho^{\text{(had)}}$ represents the resonance portion of the spectral function, $\theta$ is the Heaviside step function, and $s_0$ is the continuum threshold, and define the continuum-subtracted $0^{\text{th}}$-order LSR
\begin{equation} \label{subedLSR}
\lsr(\tau, s_0) \equiv \lsr(\tau) - \int_{s_0}^{\infty} e^{-t\tau}\frac{1}{\pi}\mathrm{Im}\Pi^{\text{(OPE)}}(t)\dif{t}\\
=\int_{t_0}^{s_0} e^{-t\tau} \rho^{\text{(had)}}(t)\dif{t}.
\end{equation}

\begin{figure}
\centering
\includegraphics[scale=0.8]{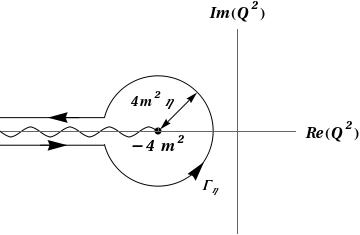}
\caption{\label{keyhole} The contour of integration used in the evaluation of the 
  LSR~(\ref{intermediateTheoryLSR})}
\end{figure}

To compute $\lsr\double{\tau}{s_0}$,
we exploit the following relation between the Borel transform and the inverse Laplace transform $\hat{\mathcal{L}}^{-1}$~\cite{Shifman:1978bx}:
\begin{equation}\label{borelIdentity}
\begin{aligned}
  \frac{1}{\tau}\hat{\mathcal{B}} \Big\{ f(Q^2) \Big\}&=\hat{\mathcal{L}}^{-1} \Big\{ f(Q^2) \Big\}\\
  &= \frac{1}{2 \pi i} \int_{c-i \infty}^{c+i \infty} f(Q^2) e^{Q^2 \tau} \dif{Q^2}
\end{aligned}
\end{equation}
where $c \in \mathbb{R}$ is selected such that $f(Q^2)$ is analytic for $\mathrm{Re}(Q^2)>c$. 
Generalized hypergeometric functions of the form $\hypgeom{p}{p-1}(z)$, (such as those appearing in $\Pi^{\text{(OPE)}}$) have a branch cut originating at the branch point $z=1$ that extends along the positive real semi-axis. As such, in the complex $Q^2$-plane, $\Pi^{\text{(OPE)}}$ is analytic everywhere except along the negative real semi-axis for $z < -Q^2/(4m^2)$. In~(\ref{borelIdentity}), we let $f \to \Pi^{\text{(OPE)}}$ and warp the contour of integration to that shown in Figure~\ref{keyhole}. We then apply definitions~(\ref{zeroThLSR}) and~(\ref{subedLSR}) to get
\begin{equation}\label{intermediateTheoryLSR}
  \lsr\double{\tau}{s_0}=\int_{4m^2(1+\eta)}^{s_0}e^{-t\tau}\frac{1}{\pi}
    \text{Im}\Pi^{\text{(OPE)}}(t)\dif{t}
  +\frac{1}{2\pi i}\int_{\Gamma_{\eta}} e^{Q^2 \tau}
    \Pi^{\text{(OPE)}}(Q^2) \dif{Q^2}\ \;\; \text{for}\ \;\; \eta\rightarrow0^{+}
\end{equation}
where
\begin{equation}\label{breakdown}
  \text{Im}\Pi^{\text{(OPE)}}(t)=
  \sum_{i=\rom{1}}^{\rom{6}}\text{Im}\Pi^{\text{(i)}}(t)
\end{equation}
and, from~(\ref{expfourd})--(\ref{expDiag6}) and~(\ref{expRenormedPert})
\begin{gather} \label{ImP}
\begin{split}
\mathrm{Im}\Pi^{\text{(\rom{1})}}(t) = 
\frac{\alpha_s}{108\pi^2 t^2 \sqrt{t-4m^2}}
\Bigg(
  12 m^2 \sqrt{t-4 m^2} \left(20 m^6-6 m^4 t-6 m^2 t^2+5 t^3\right) \sinh ^{-1}\left(\frac{1}{2m} \sqrt{t-4 m^2}\right)
  \\
  +\sqrt{t} \left(t - 4 m^2\right) \left(60 m^6+22 m^4 t-7 m^2 t^2+30 m^2 t \left(t - 4 m^2\right) \log \left(\frac{m^2}{\mu ^2}\right)-6 t^3\right)
\Bigg)
\end{split}
\\
\label{Imfourd}
  \mathrm{Im}\Pi^{\text{(\rom{2})}}(t)
  = \frac{-m^2}{6\sqrt{t(t-4m^2)}}\glueFourD
\\
\label{Imsixd}
  \mathrm{Im}\Pi^{\text{(\rom{3})}}(t)=
  \frac{24 m^6+76 m^4 t-28 m^2 t^2+3 t^3}{288 \pi t^{3/2} (t-4m^2)^{5/2}}
  \glueSixD
  \\
\label{ImDiag4}
  \mathrm{Im}\Pi^{\text{(\rom{4})}}(t)=
  \frac{6 m^4+19 m^2 t-3 t^2}{288 \pi t^{3/2} (t-4m^2)^{3/2}}
  \glueSixD
  \\
\label{ImDiag5}
  \mathrm{Im}\Pi^{\text{(\rom{5})}}(t)=
  \frac{m^2(6m^2-t)}{96 \pi t^{3/2} (t-4m^2)^{3/2}}
  \glueSixD
  \\
\label{ImDiag6}
  \mathrm{Im}\Pi^{\text{(\rom{6})}}(t)=
  \frac{- 64 \pi \alpha_s^2 (24 m^6+76 m^4 t-28 m^2 t^2+3 t^3)}{81 t^{3/2} (t-4m^2)^{5/2}}
  \big\langle \overline{q}q \big\rangle^2.
\end{gather}

Evaluating the integrals on the right-hand side of~(\ref{intermediateTheoryLSR}) for all six OPE terms leads to several divergences in $\eta$ that, when summed, delicately cancel leaving us with a finite LSR. 
A detailed treatment of the evaluation of~(\ref{intermediateTheoryLSR}) for similar inputs is available in~\cite{Palameta:2017ols}. 
Here, for the sake of brevity, we omit these details and present the LSR: 
\begin{multline}\label{finalTheoryLSR}
  \lsr\double{\tau}{s_0}=\int_{4m^2}^{s_0}e^{-t\tau}\frac{1}{\pi}
    \Bigg(\text{Im}\Pi^{\text{(\rom{1})}}(t)+\text{Im}\Pi^{\text{(\rom{2})}}(t) +p(t) \Bigg) \dif{t} \\
 -\frac{m \; e^{-4m^2 \tau}}{384\pi^2}
  \Bigg(
  \sqrt{\pi\tau} \Big( 3 - 8m^2 \tau \Big) \mathrm{erf}\Big(\sqrt{(s_0-4m^2)\tau}\Big) + \frac{162 e^{-s_0 \, \tau}}{(s_0 - 4m^2)^{3/2}} \bigg( \\
  -8 e^{s_0 \, \tau} m^2 \sqrt{\pi} \big((s_0 - 4m^2) \tau \big)^{3/2} + e^{4m^2 \, \tau} \big( 3 s_0 +32 m^4 \tau -8 m^2 (1+s_0 \tau)\big) \\
 +6 e^{s_0 \, \tau} m^2 \, \mathrm{E}_{5/2}\big( (s_0 -4m^2) \tau \big)
  \bigg)
  \Bigg)\glueSixD \\
 -\frac{8 \; m \; \alpha_{s}^{2} \; e^{-8m^2 \tau}}{243}
  \Bigg(
196 m^2 \tau ^{3/2} e^{4 m^2 \tau } \left(4 \sqrt{\pi }-3 \; \Gamma \left(-\frac{3}{2},\left(s_{0}-4 m^2\right) \tau \right)\right) + \frac{31 e^{-s_0 \, \tau}}{(s_0 - 4m^2)^{3/2}} \bigg( \\
2 e^{8 m^2 \tau } \left(32 m^4 \tau -8 m^2 (s_{0} \tau +1)+3 s_0\right) \\
-2 \sqrt{\pi \tau} \left(s_0-4 m^2\right)^{3/2} \left(8 m^2 \tau -3\right) e^{\tau  \left(4 m^2+s_0\right)} \text{erf}\left(\sqrt{\tau  \left(s_0-4 m^2\right)}\right)
  \bigg)
  \Bigg)\big\langle \overline{q}q \big\rangle^2
\end{multline}
where
\begin{multline}\label{pExplicit}
p(t)= \frac{-1}{20736 \pi  t^{3/2} \left(\sqrt{t}+2m\right)^2 \sqrt{t-4 m^2}} 
\Bigg(
81 m \left(16 m^3+16 m^2 \sqrt{t}+4 m t+t^{3/2}\right) 
\glueSixD \\
+ 2048 \pi ^2 \alpha_{s}^{2} \left(48 m^4+48 m^3 \sqrt{t}+188 m^2 t+127 m t^{3/2}+24 t^2\right) 
\big\langle \overline{q}q \big\rangle^2
\Bigg)
\end{multline}
and the imaginary parts $\mathrm{Im}\Pi^{\text{(I)}}$ and $\mathrm{Im}\Pi^{\text{(II)}}$ are given in~(\ref{ImP}) and~(\ref{Imfourd}) respectively. 
The integral on the right-hand side of~(\ref{finalTheoryLSR}) can be evaluated analytically; however, we omit the result for the sake of brevity.

Renormalization-group improvement~\cite{Narison:1981ts} requires that the strong coupling and quark mass get replaced by their corresponding running quantities evaluated at renormalization scale $\mu$. 
At one-loop in the $\overline{\text{MS}}$ renormalization scheme, for charmonium we have
\begin{gather}
  \alpha_s\rightarrow \alpha_s(\mu) = \frac{\alpha_s(M_{\tau})}{1 + \frac{25 \alpha_s(M_{\tau})}%
  {12\pi}\log\!{\Big(\frac{\mu^2}{M_{\tau}^2}\Big)}}
  \\ 
  m\rightarrow m_{c}(\mu) = \overline{m}_{c}\bigg(\frac{\alpha_s(\mu)}
  {\alpha_s(\overline{m}_{c})}\bigg)^{12/25}
\end{gather}
and for bottomonium
\begin{gather}
  \alpha_s\rightarrow \alpha_s(\mu) = \frac{\alpha_s(M_Z)}{1 + \frac{23 \alpha_s(M_Z)}%
  {12\pi}\log\!{\Big(\frac{\mu^2}{M_Z^2}\Big)}}
  \\ 
  m\rightarrow m_{b}(\mu) = \overline{m}_{b}\bigg(\frac{\alpha_s(\mu)}
  {\alpha_s(\overline{m}_{b})}\bigg)^{12/23}
\end{gather}
where~\cite{Olive:2016xmw}
\begin{gather}
  \alpha_s(M_{\tau})=0.330\pm0.014\label{alphatau}\\
  \alpha_s(M_{Z})=0.1185\pm0.0006\label{alphaZ}\\
  \overline{m}_c=(1.275\pm0.025)\ \text{GeV}\label{charmMass}\\
  \overline{m}_b=(4.18\pm 0.03)\ \text{GeV}\label{bottomMass}.
\end{gather}
For charmonium, $\mu \to \overline{m}_c$ and for bottomonium, $\mu \to \overline{m}_b$.
Finally, the following values are used for the gluon and quark
condensates~\cite{Launer:1983ib,Narison2010,ChenKleivSteeleEtAl2013}:
\begin{gather}
  \glueFourD=(0.075\pm0.02)\ \text{GeV}^4\label{glueFourDValue}\\
  \glueSixD=((8.2\pm1.0)\ \text{GeV}^2)\label{glueSixDValue}\glueFourD\\
 \big\langle \overline{q}q \big\rangle = -(0.23 \pm 0.03)^3\ \text{GeV}^3\label{quarkThreeDValue}.
\end{gather}

\section{Analysis and Results}\label{IV}
We now turn our attention to $\rho^{\text{(had)}}$ (recall~(\ref{ResCont})) which represents the resonance portion of the hadronic spectral function and contains the experimentally determined resonances we wish to probe for conventional meson-hybrid mixing. Resonances in $\rho^{\text{(had)}}$ which couple to both the conventional meson current~(\ref{CurMes}) and the hybrid current~(\ref{CurHyb}) can be thought of as meson-hybrid mixtures. 

Our analysis approach is to build a variety of models of the $1^{++}$ heavy quarkonium mass spectra (i.e. a variety of choices for $\rho^{\text{(had)}}$) that take known resonance masses as inputs, and test them for conventional meson-hybrid mixing. 
In Table~\ref{charmoniumTable}, we list all $1^{++}$ charmonium resonances that have a Particle Data Group entry in~\cite{Olive:2016xmw}, and in Table~\ref{bottomoniumTable}, we do the same for bottomonium. Note that, in Table~\ref{charmoniumTable} and Table~\ref{bottomoniumTable}, all entries have $I^G=0^{+}$.

Laplace sum-rules are generally insensitive to resonance widths, and so we consider $\rho^{\text{(had)}}$ to be a sum of narrow resonances, i.e.,
\begin{equation}\label{rhoForms}
  \rho^{\text{(had)}}(t)=\sum_{i=1}^n \xi_i \delta(t-m_i^2)
\end{equation}
where $n$ is the number of resonances in the model. 
The $\{\xi_i\}_{i=1}^n$ are mixing parameters 
(products of hadron masses, signed hadronic couplings, and mixing angle factors)
which are a measure of the combined coupling to both the conventional meson current and hybrid current. 
For example, in a simple case of two-state mixing, we would have 
$\xi_1=m_H^2 m_M^2 f_H f_M \sin^2\theta$ and 
$\xi_2=-m_H^2 m_M^2 f_H f_M \cos^2\theta$ where $\theta$ is a mixing angle between pure hybrid and meson states with corresponding couplings $f_H$ and $f_M$.
A state with both conventional meson and hybrid components has $\xi_i\neq0$. A pure conventional meson state or pure hybrid state has $\xi_i=0$. The specific models for which we present results are given for the charmonium and bottomonium sectors in Tables~\ref{charmoniumModels} and~\ref{bottomoniumModels} respectively.

\begin{table}
\caption{Particle Data Group masses of $1^{++}$ charmonium resonances~\cite{Olive:2016xmw}.}
\label{charmoniumTable}
\centering
\begin{tabular}{lS}
  \addlinespace
  \toprule
  Name &  \multicolumn{1}{c}{Mass (\si{GeV})} \\
  \midrule
  $\chi_{c1}(1P)$ & 3.51 \\
  $X(3872)$ & 3.87 \\
  $X(4140)$ & 4.15 \\
  $X(4274)$ & 4.27 \\
  \bottomrule
\end{tabular}
\end{table}
\begin{table}
\caption{Particle Data Group masses of $1^{++}$ bottomonium resonances~\cite{Olive:2016xmw}.}
\label{bottomoniumTable}
\centering
\begin{tabular}{lS}
  \addlinespace
  \toprule
  Name &  \multicolumn{1}{c}{Mass (\si{GeV})} \\
  \midrule
  $\chi_{b1}(1P)$ & 9.89 \\
  $\chi_{b1}(2P)$ & 10.26 \\
  $\chi_{b1}(3P)$ & 10.51 \\
  \bottomrule
\end{tabular}
\end{table}

\begin{table}
\centering
\caption{A representative collection of hadron models analyzed in the charmonium sector.}
\label{charmoniumModels}
\begin{tabular}{ccccc}
  \addlinespace
  \toprule
  Model & $m_1$ & $m_2$ & $m_3$ & $m_4$\\
   & (\si{GeV}) & (GeV) & (GeV) & (GeV)\\ 
  \midrule
  C1 & 3.51 & - & - & -  \\
  C2 & 3.51 & 3.87 & - & -  \\
  C3 & 3.51 & 3.87 & 4.15 & -  \\
  C4 & 3.51 & 3.87 & 4.15 & 4.27  \\
  \bottomrule
\end{tabular}
\end{table}

\begin{table}
\centering
\caption{A representative collection of hadron models analyzed in the bottomonium sector.}
\label{bottomoniumModels}
\begin{tabular}{cccc}
  \addlinespace
  \toprule
  Model & $m_1$ & $m_2$ & $m_3$ \\
   & (GeV) & (GeV) & (GeV)\\ 
  \midrule
  B1 & 9.89 & - & - \\
  B2 & 9.89 & 10.26 & - \\
  B3 & 9.89 & 10.26 & 10.51 \\
  \bottomrule
\end{tabular}
\end{table}

Substituting~(\ref{rhoForms}) into~(\ref{subedLSR}) gives
\begin{equation}\label{lsrFinal}
  \lsr\double{\tau}{s_0}=\sum_{i=1}^n \xi_i e^{-m_i^2 \tau}.
\end{equation}
To extract hadronic properties from~(\ref{lsrFinal}) together with LSR~(\ref{finalTheoryLSR}),
we must first, for each model, select an acceptable range of $\tau$ values, i.e., a Borel window $\double{\tau_{\text{min}}}{\tau_{\text{max}}}$. 
To determine the Borel window, we follow the same methodology as in~\cite{Chen:2013pya,Ho:2016owu,BergHarnettKleivEtAl2012,Harnett:2012gs}.
To select $\tau_{\text{min}}$, we consider
\begin{equation}\label{pole_contribution}
  \frac{\lsr\double{\tau}{s_0}}{\lsr\double{\tau}{\infty}},
\end{equation}
i.e., the ratio of the LSR's hadron contribution to its hadron plus continuum contribution. We demand that this ratio be at least 10\%. To select $\tau_{\text{max}}$, we demand that the LSR converge where convergence is taken to mean that the magnitude of successive OPE terms be at most one-third that of any previous term. This means that we require the magnitude of the 4d gluon condensate contribution be less than one-third that of the perturbative contribution. We also require that the magnitude of the sum of the 6d gluon and quark condensate contributions be less than one-third that of 4d gluon condensate contribution. 

For particular choices of $\{m_i\}_{i=1}^n$, the quantities $\{\xi_i\}_{i=1}^n$ and $s_0$ are extracted as best fit parameters to~(\ref{lsrFinal}). To do so, we partition the Borel window into $N=20$ equal length subintervals with $\{\tau_j\}_{j=0}^N$, and define
\begin{equation}\label{chiSquaredGeneral}
  \chi^2(\xi_1,\ldots,\xi_n,\,s_0)=\sum_{j=0}^N
  \Bigg(
    \lsr\double{\tau_j}{s_0}-
    \sum_{i=1}^n \xi_i e^{-m_i^2 \tau_{j}}
  \Bigg)^2.
\end{equation}
Minimizing~(\ref{chiSquaredGeneral}) gives predictions for $\{\xi_i\}_{i=1}^n$ and $s_0$ corresponding to the best fit agreement between QCD and the hadronic model in question.

The procedure described above for selecting a Borel window depends on $s_0$.
However, $s_0$ is not known at the outset.
It is one of the parameters that emerges from the minimization of~(\ref{chiSquaredGeneral}).
But the definition of~(\ref{chiSquaredGeneral}) requires a Borel window.
Hence, we determine both the Borel window and $s0$ iteratively.
We start with a seed value of $s_0 = 2m_{\text{max}}^{2}$ where $m_{\text{max}}$ is the mass of the heaviest resonance in the model. This seed value separates the continuum from the resonances by a generous margin. We generate a Borel window for this $s_0$ value according to the criteria outlined above. Minimization of~(\ref{chiSquaredGeneral}) then yields an updated value for $s_0$. This process is iteratively repeated until $s_0$ and the Borel window settle. 
For all the models examined in the charmonium sector, we found that the Borel window settled to 
$\tau_{\text{min}}=0.17\ \text{GeV}^{-2}$ to $\tau_{\text{max}}=0.41\ \text{GeV}^{-2}$,
and, in the bottomonium sector, all of the models have Borel windows that settled to $\tau_{\text{min}}=0.02\ \text{GeV}^{-2}$ to $\tau_{\text{max}}=0.12\ \text{GeV}^{-2}$. 
These persistent values for the Borel window across different models in each sector demonstrates the LSR's insensitivity to changes in $s_0$ and is consistent with our findings in~\cite{Palameta:2017ols}. 

\begin{figure}
\centering
\includegraphics[scale=0.7]{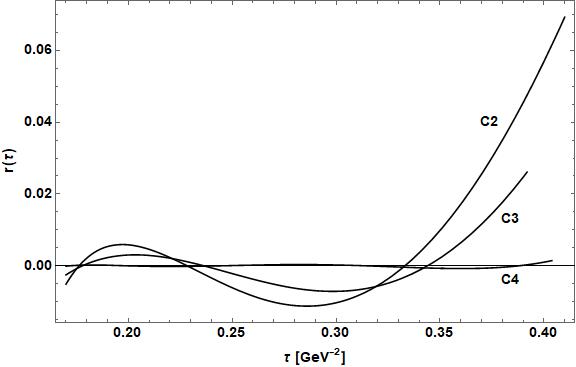}
\caption{\label{residualsC} Relative residuals~(\ref{residuals})
for models C2--C4 using the optimized values of $s_0$ and $\xi_i$
from Table~\ref{charmoniumResults}. Residuals for model C1 are not shown because they are much larger than  the scales of the figure.}
\end{figure}

\begin{figure}
\centering
\includegraphics[scale=0.7]{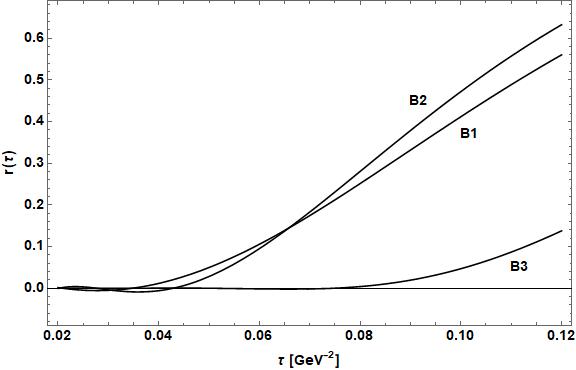}
\caption{\label{residualsB} Relative residuals~(\ref{residuals})
for models B1--B3 using the optimized values of $s_0$ and $\xi_i$
from Table~\ref{bottomoniumResults}.
}
\end{figure}

We extract $\{\xi_i\}_{i=1}^n$ and $s_0$ for each of the models defined in Tables~\ref{charmoniumModels} and~\ref{bottomoniumModels}, and present our results in Tables~\ref{charmoniumResults} and~\ref{bottomoniumResults} respectively. 
Instead of presenting each $\xi_i$, we present $\zeta$ and $\frac{\xi_i}{\zeta}$ where
\begin{equation}\label{zeta}
  \zeta=\sum_{i=1}^n |\xi_i|.
\end{equation}
The errors included are associated with the strong coupling values~(\ref{alphatau})--(\ref{alphaZ}), the quark masses~(\ref{charmMass})--(\ref{bottomMass}), the condensates~(\ref{glueFourDValue})--(\ref{quarkThreeDValue}), and an allowed $\pm0.1$~GeV variability in the renormalization scale~\cite{Narison:2014ska}. We also allow for the end points of the Borel window to vary by $0.1\ \text{GeV}^{-2}$ in the charmonium sector and $0.01\ \text{GeV}^{-2}$ in the bottomonium sector. The vacuum saturation parameter $\kappa$ from~(\ref{sixDQcond}) is not varied because the numerical contribution to the LSR~(\ref{finalTheoryLSR}) stemming from the 6d quark condensate diagram is negligible. Our results are most sensitive to varying the value of $\tau_{\text{min}}$ and varying the value of the quark masses~(\ref{charmMass}) and~(\ref{bottomMass}).
In Figure~\ref{residualsC}, we plot relative residuals
representing the difference between the QCD prediction and the resonance plus continuum hadronic model,
\begin{equation}\label{residuals}
r(\tau)= \frac{\lsr\double{\tau}{s_0}-\sum_{i=1}^n \xi_i e^{-m_i^2 \tau}}{\lsr\double{\tau}{s_0}},
\end{equation}
(the numerator in~(\ref{residuals}) is
the difference between the left- and right-hand sides
of~(\ref{lsrFinal})) for models~C2--C4 using the optimized values of $s_0$ and $\{\xi_i\}$
from Table~\ref{charmoniumResults}.  
In Figure~\ref{residualsB}, we do the same for models~B1--B3 using the optimized values from
Table~\ref{bottomoniumResults}.

\begin{table}
\centering
\caption{Continuum thresholds and $\chi^2$ values for hadron models defined in Table~\ref{charmoniumModels} and their resulting extracted mixing parameters with their theoretical uncertainties.}
\label{charmoniumResults}
\begin{tabular}{cSSSSSSS}
\addlinespace
\toprule
  Model & \multicolumn{1}{c}{$s_0$} & {$\chi^2\times10^9$} 
    & \multicolumn{1}{c}{$\zeta$} 
    & {$\frac{\xi_1}{\zeta}$} 
    & \multicolumn{1}{c}{$\frac{\xi_2}{\zeta}$} 
    & \multicolumn{1}{c}{$\frac{\xi_3}{\zeta}$}
    & \multicolumn{1}{c}{$\frac{\xi_4}{\zeta}$}\\
   & {($\text{GeV}^2$)} & {($\text{GeV}^{12}$)} & {($\text{GeV}^{6}$)} & & &\\
\midrule
  C1 & 18.8 & 7990 & 0.18 \pm 0.01 & 1 & {-} & {-} & \multicolumn{1}{c}{-} \\
  C2 & 28.8 & 76.3 & 0.83 \pm 0.07 & 0.47 \pm 0.02 & -0.53 \pm 0.02 & {-} & \multicolumn{1}{c}{-} \\
  C3 & 18.8 & 27.4 & 2.6 \pm 0.4 & 0.21 \pm 0.02 & -0.45 \pm 0.01 & 0.34 \pm 0.02 & {-} \\
  C4 & 31.7 & 0.0586 & 44 \pm 6 & 0.03 \pm 0.01 & -0.16 \pm 0.01 & 0.46 \pm 0.01 & -0.35 \pm 0.01 \\
\bottomrule
\end{tabular}
\end{table}

\begin{table}
\centering
\caption{Continuum thresholds and $\chi^2$ values for hadron models defined in Table~\ref{bottomoniumModels} and their resulting extracted mixing parameters with their theoretical uncertainties.}
\label{bottomoniumResults}
\begin{tabular}{cSSSSSS}
\addlinespace
\toprule
  Model & {$s_0$} & {$\chi^2 \times 10^6$} 
  & {$\zeta$} 
  & {$\frac{\xi_1}{\zeta}$} 
  & {$\frac{\xi_2}{\zeta}$}
  & {$\frac{\xi_3}{\zeta}$} \\
   & {($\text{GeV}^2$)} & {($\text{GeV}^{12}$)} & {($\text{GeV}^{6}$)} & & &\\
\midrule
  B1 & 128 & 2580 & 49 \pm 1 & 1 & {-} & {-} \\
  B2 & 282 & 1980 & 70 \pm 4 & 0.30 \pm 0.01 & 0.70 \pm 0.01 & {-} \\
  B3 & 241 & 0.832 & 1905 \pm 28 & 0.16 \pm 0.01 & -0.48 \pm 0.01 & 0.36 \pm 0.01 \\
\bottomrule
\end{tabular}
\end{table}

\section{Discussion}\label{V}
As shown in Tables \ref{charmoniumResults} and \ref{bottomoniumResults}, the inclusion of heavy resonances beyond the ground state 
significantly improves agreement between QCD and experiment 
in both the charmonium and bottomonium sectors. 
In particular, in the charmonium sector, going from three to four resonances 
(i.e., from model~C3 to~C4 in Table~\ref{charmoniumResults}) 
decreases the value of the $\chi^2$ (recall~(\ref{chiSquaredGeneral})) by a factor of 468 
while in the bottomonium sector, going from two to three resonances 
(i.e., from model~B2 to~B3 in Table~\ref{bottomoniumResults}) 
decreases the $\chi^2$ by a factor of 2380.
This improvement can also be seen from the trend of  
decreasing magnitude of relative residuals with increasing number of resonances
depicted in Figures~\ref{residualsC} and~\ref{residualsB}.
For the highest mass resonance in a given model we define a measure of its contribution to the LSR as in~\cite{Palameta:2017ols}:
\begin{equation}\label{signalStrength}
  \frac{\left|\int_{\tau_{\text{min}}}^{\tau_{\text{max}}} \xi_n e^{-m_n^2 \tau} d\tau\right|}%
  {\sum_{i=1}^{n}\left| \int_{\tau_{\text{min}}}^{\tau_{\text{max}}} \xi_i e^{-m_i^2 \tau} d\tau \right|}
\end{equation}
where $n$ is the number of resonances in the model. 
The highest mass resonances make substantial contributions to the LSRs in spite of the exponential suppression inherent in LSRs: 
in the charmonium sector, evaluating~(\ref{signalStrength}) for model~C4 
gives~0.25, and in the bottomonium sector, evaluating~(\ref{signalStrength}) for model~B3 gives~0.30. 
These results, coupled with the dramatic improvement in $\chi^2$-values when compared 
to models containing less resonances, indicate the significant impact that the 
highest mass resonances have on the LSRs, and cause us to favour models~C4 and~B3.

In the charmonium sector, model~C4 indicates that
there is almost no conventional meson-hybrid mixing in the $\chi_{c1}(1P)$, 
minimal mixing in the $X(3872)$,
and significant mixing in both the $X(4140)$ and $X(4274)$.  
Assuming the $\chi_{c1}(1P)$, the lightest known resonance in this sector,
has a large conventional meson component~\cite{Godfrey:1985xj}, 
then our results indicate that it has very little hybrid component.
Regarding the interpretation of the $X(3872)$, 
if the $X(3872)$ does have a significant hybrid component,
a possibility put forth in~\cite{ChenJinKleivEtAl2013}, 
then our results indicate that it does not have a significant conventional meson component.  
However, if the $X(3872)$ does have a large conventional meson component 
as argued in~\cite{Matheus:2009vq}, 
then our results indicate that it does not have a large hybrid component.
In addition, our results imply that the $X(4140)$ and the $X(4274)$ 
both contain significant conventional meson and hybrid components.

In the bottomonium sector, model~B3 indicates that
there is minimal conventional meson-hybrid mixing in the $\chi_{b1}(1P)$ and significant mixing in both
the $\chi_{b1}(2P)$ and the $\chi_{b1}(3P)$.
Thus, assuming the $\chi_{b1}(1P)$, the lightest observed resonance in this sector, 
contains a significant conventional meson component~\cite{Godfrey:1985xj},
our results imply that it does not have a large hybrid component.  
Also, our results indicate that the 
$\chi_{b1}(2P)$ and the $\chi_{b1}(3P)$ each contain significant 
conventional meson and hybrid components.

\clearpage
\section*{Acknowledgements}
We are grateful for financial support from the National Sciences and 
Engineering Research Council of Canada (NSERC).

\bibliographystyle{h-physrev}
\bibliography{research}

\end{document}